# First Measurement of a 127 keV Neutron Field with a μ-TPC Spectrometer

D. Maire, J. Billard, G. Bosson, O. Bourrion, O. Guillaudin, J. Lamblin, L. Lebreton, F. Mayet,
J. Médard, J.F. Muraz, M. Petit, J.P. Richer, Q. Riffard, and D. Santos

*Abstract*—In order to measure the energy of neutron fields, with energy ranging from 8 keV to 1 MeV, a new primary standard is being developed at the IRSN (Institute for Radioprotection and Nuclear Safety). This project, μ-TPC (Micro Time Projection Chamber), carried out in collaboration with the LPSC (Laboratoire de Physique Subatomique et de Cosmologie), is based on the nuclear recoil detector principle.

The instrument is presented with the associated method to measure the neutron energy. This article emphasizes the proton energy calibration procedure and energy measurements of a neutron field produced at 127 keV with the IRSN facility AMANDE.

## I. INTRODUCTION

In the field of ionizing radiation, facilities producing neutron fields are essential to study and to calibrate neutron detectors. To do so, neutron fields are characterized in energy and fluence by a standard spectrometer and then can be considered as reference fields. One of the IRSN facilities, called AMANDE (Accelerator for Metrology and Neutron Applications in External Dosimetry) produces mono-energetic neutron fields with an energy ranging between 2 keV and 20 MeV dedicated to neutron standard references [1]. To measure directly the energy distribution of neutron fields with energies below a few tens of keV, a new gaseous detector (μ-TPC) is being developed at the Laboratory of Metrology and Neutron Dosimetry (IRSN/LMDN). The IRSN is associated with the LNE (Laboratoire National de métrologie et d'Essais) for the French neutron references.

The measurement strategy of this μ-TPC detector is based on 3D track reconstruction of nuclear recoils down to a few keV. The 3D track reconstruction of nuclear recoils can be achieved with micro-patterned detectors, such as a gas electron multiplier (GEM) [2], a multi-wire proportional chamber (MWPC) [3] or a micromegas [4]. This μ-TPC is a low-pressure detector (50 mbar) using a micromegas coupled to a self-triggered electronics [13,14].

## II. PROJECT CONTEXT

This project is undertaken in collaboration with the MIMAC team (LPSC/UJF/CNRS-IN2P3/INP), which has developed the first MIMAC prototype [5] for directional dark matter search [6]. The direct detection of dark matter is in principle similar to the fast neutron detection in the keV range because the interaction with matter of these particles induces in both cases nuclear recoils in the same range of energy.

### A. Metrological Issues

The French references held by IRSN are dedicated to calibrate neutron detectors like dosimeters used for radiation safety of workers. Indeed accurate dose measurements are necessary for workers in nuclear industry or healthcare field as well as for medical treatments. In order to calculate the neutron equivalent dose, conversion coefficients depending on the neutron energy are applied to the measured neutron fluence. Hence, the fluence have to be measured as a function of neutron energy to calculate precisely the neutron equivalent dose delivered. To improve the quality of measurements and to be independent of other laboratories, the neutron fields have to be characterized by a primary measurement standard, i.e. the highest metrological level. This requires that the measurement procedure used to obtain the measurand (fluence or energy) must be unrelated to a measurement standard of the same kind [7]. For example, a primary measurement standard devoted to measure energy and fluence of neutron fields is therefore not calibrated using a neutron field.

At IRSN, several facilities are able to deliver a precise neutron equivalent dose. These facilities were developed as recommended by the ISO standards 8529-1 to become the French references [8][9][10][11].

### B. The AMANDE Facility

The AMANDE accelerator is a 2 MV Tandetron accelerator system providing protons and deuterons with an energy ranging between 100 keV and 4 MeV, in a continuous or in a pulsed mode. Neutrons are produced by the interaction

This paper is submitted for review the 31th may 2013. This work was supported in part by the Laboratoire National de métrologie et d'Essais (LNE) under Grant LNE/DRST/12 7 004.

D. MAIRE L. Lebreton and M. Petit are with the Institute for Radioprotection and Nuclear Safety (IRSN), site of Cadarache, 13115 Saint Paul lez Durance, FRANCE (e-mail: donovan.maire@irsn.fr).

J. Billard, was with the Laboratoire de Physique Subatomique et de Cosmologie (LPSC/CNRS-IN2P3/UJF/INP), 38000 Grenoble, France. He is now with the department of Physics, Massachusetts Institute of Technology, Cambridge, MA 02139, USA.

O. Bourrion, G. Bosson, O. Guillaudin, J. Lamblin, F. Mayet, J. Médard, J-F. Muraz, J.P. Richer, Q. Riffard, and D. Santos are with the Laboratoire de Physique Subatomique et de Cosmologie (LPSCCNRS-IN2P3/UJF/INP), 38000 Grenoble, FRANCE (e-mail: daniel.santos@lpsc.in2p3.fr).

between the ion beam and thin targets, e.g. lithium or scandium, placed at the end of the beam line. The mono-energetic neutron fields produced by AMANDE below 1 MeV and recommended by the ISO 8529-1 [8] have been detailed in [11]. The neutron energy depends on the energy of an incident charged beam (e.g. protons) and the choice of the target. This neutron energy varies also with the angle relative to the beam direction. The neutron energy spread depends on the target thickness. For example, the reference neutron energy is calculated thanks to the kinematics of the reaction and the known target thickness when the lithium target is used.

The spectrometers, used to provide neutron references, were calibrated at the Physikalisch Technische Bundesanstalt (PTB), which implies they are secondary standards. In addition the lowest detection limit is 60 keV whereas neutrons may be produced with energies down to 2 keV.

The aim of a national standard laboratory is to be independent to other national laboratories. In this context it is essential to develop instruments using a primary measurement process to determine a primary value.

### III. TECHNICAL DESCRIPTION OF THE μ-TPC

A nuclear recoil detector uses a converter to produce recoils of nuclei with mass $m_A$ thanks to the elastic scattering of neutrons from these nuclei. The nucleus energy ($E_A$) measurement and the reconstruction of the initial nucleus recoil angle ($\theta_A$), between the initial nucleus direction and the incidence of the neutron, allow reconstructing the neutron energy ($E_n$), following the equation 1:

$$E_n = \frac{(m_n + m_A)^2}{4 m_n m_A} \times \frac{E_A}{\cos^2(\theta_A)} \quad (1)$$

where $m_n$ is the neutron mass. If the recoil nucleus is a proton, $m_n \approx m_A$ which implies the factor $\frac{(m_n + m_A)^2}{4 m_n m_A}$ is approximately equal to 1. The maximum energy achievable for a nuclear recoil will be obtained on protons.

From the proton energy measurement and the reconstruction of the 3D proton track, the neutron energy can be directly calculated. This would allow the proton recoil detector to be a primary standard. The status of primary standard could be reachable for the μ-TPC described in this paper.

The μ-TPC may characterize low energy neutron fields [12], between 8 keV and 1 MeV. The use of a gas as a converter and the detection of proton recoils are the only answer to reach such an energy range.

The μ-TPC is divided in two zones with the anode having an active area of 10.8x10.8 cm$^2$: the conversion zone 17.7 cm in length and the amplification one, 256 μm in length of a bulk micromegas [4]. In the first zone, proton recoils stemming from the neutron scattering lose a part of their kinetic energy by ionizing the gas producing a number of ion-electron pairs. A field cage surrounding the conversion zone produces a uniform electric field, enabling the electrons coming from the ionization to drift toward the amplification zone. In the amplification zone a second field, of a much higher value, produces an avalanche, which amplifies the signal up to the pixelised anode. The ions produced in this avalanche drift back to the grid and the anode collects the electrons.

The signal measured on the grid is amplified by a charge sensitive preamplifier and sent to the flash ADC sampled at 50 MHz. The measured ionization energy is calculated as the difference between the maximum amplitude and the minimum of the ADC signal. The amplification voltage and the gas pressure can be adjusted to measure a large range of neutron energies.

The anode is segmented in pixels with a pitch of 424 μm. Reading 256 strips in each dimension to access the X and Y positions perform the 2D readout of the anode. The pixelised anode is entirely read with a frequency of 50 MHz thanks to a self-triggered electronics developed at the LPSC. This electronics is composed of 8 ASICs with 64 channels each, associated with a data acquisition system [13][14]. This electronics allows to sample in two dimensions the proton track every 20 ns. The third dimension is therefore reconstructed by using the drift velocity of electrons in the conversion zone. Each mixture gas has its own drift velocity depending on the electric field and pressure. The drift velocity may be measured independently by using an alpha source as was shown in [19]. The alpha particles crossing the whole chamber ionize the gas between the cathode and the anode. The drift time of electrons, defined as the time between the first electrons collected and the last ones taking into account the convolution of the electronics response [15][16], allows calculating the drift velocity of electrons in the experimental conditions. Measurements of the electron drift velocity match rather well with the Monte Carlo code MAGBOLTZ [17]. Therefore MAGBOLTZ was used to evaluate the electron drift velocity for this experiment. The angle of the proton track, with respect to the beam axis, is deduced from this 3D reconstruction.

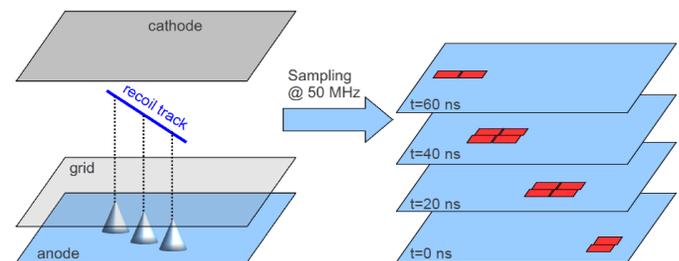

Fig. 1. Detection principle of the μ-TPC. Proton recoils coming from n-p elastic scattering ionize the gas. The charge cloud is amplified and electrons are collected on the pixelised anode. The anode sampling allows the 3D reconstruction of the track.

The collection of charges on the grid enables the measurement of the proton energy lost by ionizing the gas and eventually the initial proton energy.

The gas mixture used is: 60% $C_4H_{10}$ and 40% $CHF_3$ at 50 mbar. The $C_4H_{10}$ was chosen due to the high proportion of hydrogen, which increases the efficiency of the detector. The $C_4H_{10}$ is in addition a good quencher due to its many vibrational and rotational states that allows the absorption of low energy photons produced in the avalanches, in the range of few eV coming from the de-excitation of molecular states.

The CHF$_3$ allows lowering the drift velocity to obtain more images of the tracks with the same sampling frequency (50 MHz). The gas flow is provided by a gas control system dedicated to this detector. This system enables the pressure and the composition of the gas to be changed in order to adapt the converter to the neutron energies. Each gas is filtered to remove impurities such as $O_2$ and $H_2O$ molecules.

A 3D projection of a recoil proton track at an ionization energy of 100 keV, measured during this experiment, is shown in the figure 2 associated with its flash ADC profile. The XY plane represents the 256 x 256 pixels anode.

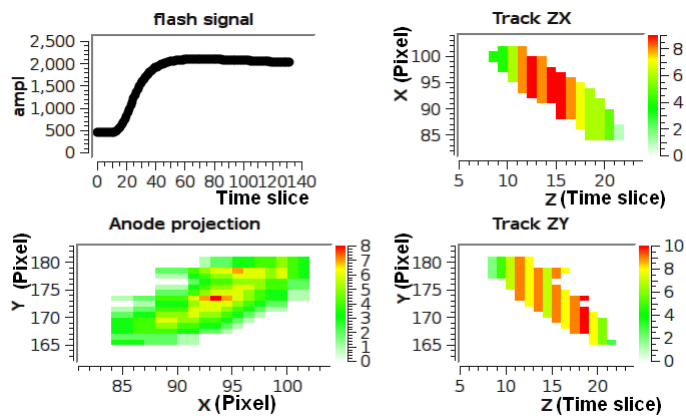

Fig. 2. Detection of a 100 keV recoil nuclei compatible with a proton recoil. The plot on the upper left is the flash ADC profile due to the charge collection on the grid as a function of the time in 20 ns units. The particle ionization energy is defined as the difference between the maximum and the minimum of the amplitude of the signal. The other plots correspond to the projection recoil track on the anode plane (XY plane), on the ZX plane and ZY plane. The Z coordinate corresponds to the time sampling (i.e. 20 ns for each time slice).

## IV. NEUTRON ENERGY MEASUREMENTS

The neutron energy is inferred from the measurement of the initial recoil angle, $\theta_P$, and the initial proton energy, $E_p$ (equation 1). Methods to measure both quantities are discussed in the following sections.

### A. Proton Energy Calibration Principle

To get the energy calibration, two X ray sources are used in the chamber. The $^{109}$Cd produces L-shell X rays with a mean energy of 3.04 keV and K-shell X rays with a mean energy of 22.1 keV. In addition K-shell X rays, produced by a source of $^{55}$Fe, with a mean energy of 5.96 keV are measured. The X rays interact by photoelectric effect in the detector. The photoelectrons lose their kinetic energy by ionizing the gas and the electrons are collected on the anode. Due to the transparency of the gas at such low pressure to high X ray energies and the fact that photoelectrons have to remain totally in the active detection zone, only X rays with energies lower than 10 keV can be used for the calibration. This calibration is processed just before and just after the measurements for monitoring probable variation of the gain during the experiment.

The amplification voltage was adjusted to get the maximum nuclear recoil energy, i.e. the endpoint, well below the highest ADC channel (4096). The calibration spectrum obtained is shown in the figure 3.

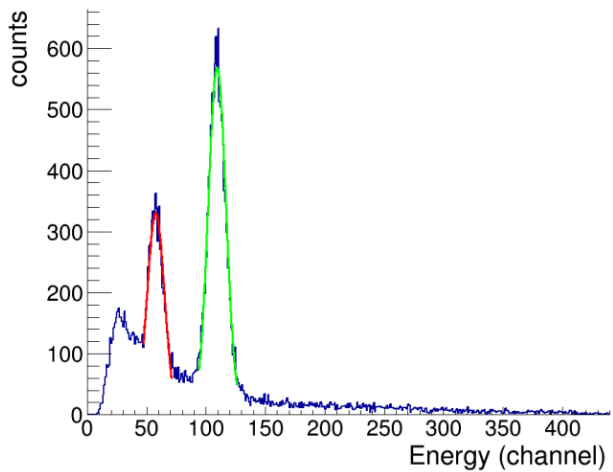

Fig. 3. Energy calibration with two X ray sources: $^{109}$Cd (3.04 keV) and $^{55}$Fe (5.96 keV). The gas mixture used is $C_4H_{10}$ with 40% CHF$_3$ at 50 mbar. Two peaks are fitted with a Gaussian function to obtain the mean energy value and the associated resolution (FWHM). The background is mainly due to an incomplete charge collection. The long tail at the right of the peaks shows the contribution of the incomplete charge collection of the photoelectron tracks of 22.1 keV (K-shell X-rays produced by the decay of $^{109}$Cd) due to the finite size of the detector and their Compton electrons contribution.

The figure 3 shows two peaks well separated corresponding to 3.04 keV (red curve) and 5.96 keV (green curve). The resolution, defined as the Full Width at Half Maximum (FWHM) over the mean, is 29% and 16% respectively for these two peaks. Both peaks are fitted by a Gaussian function to get a linear calibration.

The continuum observed in the energy distribution is coming from the incomplete collection of electrons with energies greater than 10 keV and their Compton electron contribution. Photoelectron tracks, with energies of 22.1 keV, are not detected entirely due to the finite size of the detector.

The energies used to calibrate the detector are relatively low compared to the maximum proton recoil energy produced (i.e. 127 keV). In addition, only two energies are used and their uncertainty is 0.2 keV for the $^{109}$Cd source and 0.04 keV for the $^{55}$Fe source according to the database ENDF B-VII.1. This linear calibration induced by extrapolation a maximum probable uncertainty of 12 keV at 127 keV energy. In order to improve the calibration in the range of 50-100 keV along with a measurement of the quenching factor in ionization of protons a dedicated device has been developed at LPSC to produce electrons and protons up to 50 keV coupled to an ionization chamber with the same gas used in our detector. This device, called COMIMAC shown in fig. 4, is a portable quenching equipment that will be described in a future paper. The COMIMAC device allows producing ion beams with masses up to A=20, with a defined energy from 1 to 50 keV as well as electrons in the same range of energies becoming an ideal tool to measure the IQF of such nuclei.

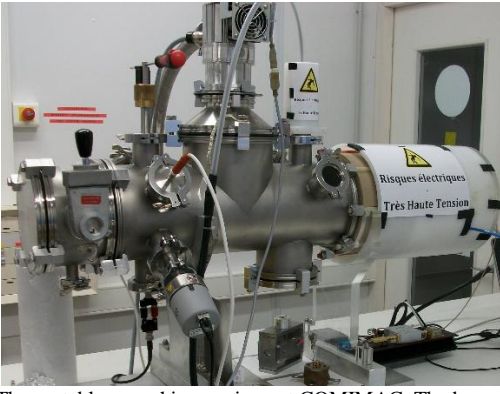

Fig. 4. The portable quenching equipment COMIMAC. The beam is extracted from a plasma produced in a COMIC source [22] on the right part, and then sent to the ionization chamber on the left.

Protons lose only a part of their kinetic energy by ionizing the gas. In this way the measured energy is only the ionization energy, $E_{ion}$. To measure the initial proton energy the Ionization Quenching Factor (IQF) has to be estimated. This factor is defined by the yield between the measured ionization energy of the nuclear recoil and the ionization energy of the electronic recoil with the same initial energy, $E^e_{initial}$ (equation 2) [16].

$$Q = \frac{E_{ion}}{E^e_{initial}} \quad (2)$$

This factor depends on the gas mixture and the initial proton energy. The knowledge of this factor is required to measure low proton energy (i.e. $E_{initial} < 50$ keV) because the fraction of energy lost by ionization decreases exponentially with proton energy.

The IQF has never been measured for protons at such energies in a gas mixture of $C_4H_{10}$ and $CHF_3$. Therefore this factor is calculated for this analysis with SRIM [18] in the gas mixture 60% $C_4H_{10}$ and 40% $CHF_3$ at 50 mbar. The result of this simulation is shown in the figure 5.

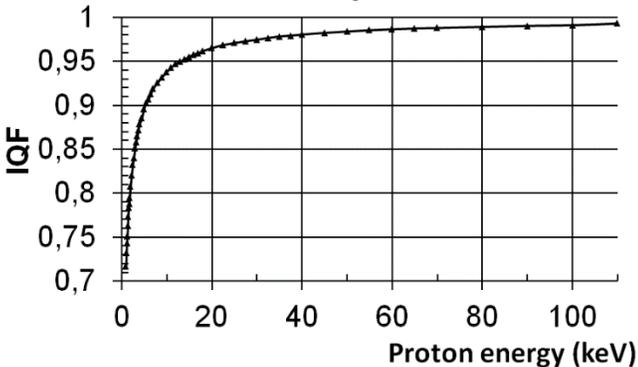

Fig. 5. Proton Ionization Quenching Factor derived from SRIM calculation of the proton stopping power. The gas mixture used is $C_4H_{10}$ with 40% $CHF_3$ at 50 mbar.

The IQF decreases with decreasing proton energy as expected. But previous studies, performed by the MIMAC team with alpha particles in $^4He + 5\%$ $C_4H_{10}$ at 350 mbar [19], have shown that SRIM calculations overestimate the IQF up to 20% of the total kinetic energy.

SRIM does not recreate the real experimental conditions: pollutions (e.g. $O_2$ or $H_2O$) in the gas mixture or little changes of the gas composition. The proton IQF is probably overestimated for this analysis using SRIM, which implies the measured neutron energy, is probably lower than the produced neutron energies. The proton IQF has to be measured to improve the estimation of the proton energy and to reduce uncertainties on the measurement of the neutron energies. The COMIMAC device will help to improve this important point allowing measurements of the IQF for the low energy protons (i.e. $E_p < 50$ keV).

### B. Neutron Energy Measurements on AMANDE

The measurement campaign presented in this paper was carried out in March 2013, and was dedicated to test the reconstruction process with a real neutron field.

Although the neutron energy 144 keV is recommended by the ISO 8529-1, measurements were performed on a neutron field with energy of 127 keV for this experimental campaign. The detector is mainly made of Aluminum and the scattering cross section of Al has a resonance at 144 keV. Therefore the neutron energy was lowered to reduce the neutron scattering in the detector walls. The scattering cross section of Al, given by ENDF B-VII.1, decreases from 20 to 2 barns for 144 keV and 127 keV respectively.

The Z-axis of the μ-TPC was set at 0° with respect to the beam line. The distance between the LiF target and the front side of the μ-TPC was fixed at 72.5 cm.

The acquisition system can be operated in two modes: i) only the grid signal giving the ionization energy is required, ii) the grid signal and a coincidence between X and Y strips of pixels are required. Each event is always associated to one ionization energy. In addition it may fire in coincidence X and Y strips of pixels. The addition of $CHF_3$ to the gas mixture reduce the drift velocity of the electrons producing a dilution of the charge per unit of time sampling preventing the strips of pixels to be fired with electron tracks. The electron tracks are one order of magnitude longer than the recoil tracks of the same ionization energy. In consequence the coincidence mode enables to have a first discrimination of nuclear tracks from photoelectrons, produced by gamma or X rays. The figure 6 shows the ionization energy distribution obtained from the flash ADC signal on the grid, in ADC units, during neutron energy measurements. The black curve is the spectrum measured without the coincidence mode. The gray curve, with the area under the curve colored in pink, is obtained with the coincidence mode. This distribution represents the ionization energy distribution of mainly nuclear recoils. This distribution is the typical flat distribution expected with a mono-energetic neutron source. To validate that photoelectrons were mainly removed from the spectrum with the coincidence mode, measurements were performed with the backing of the target, made of Tantalum and Fluorine, producing only photons. The coincidence enabled to remove more than 99% of events when we use only the backing as target.

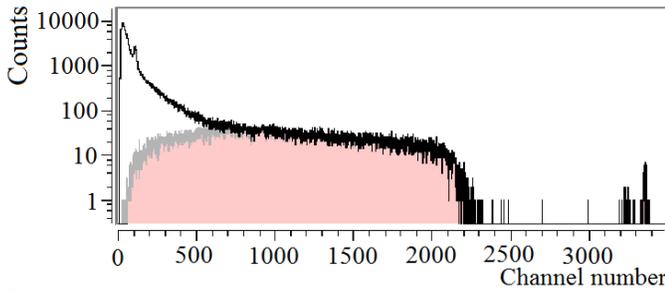

Fig. 6. Ionization energy spectra from the flash ADC obtained by the charge collection on the grid (black curve) when a neutron field is produced by AMANDE. The gray curve (pink area) is the spectrum obtained when a coincidence between strips of pixels and the grid is required (coincidence mode). The peak due to the $^{55}$Fe source is still visible at channel 109 on the black distribution. Alpha particles emitted during the experiment saturate the flash ADC around the channel 3300.

The peak due to the $^{55}$Fe source is still visible at channel 109 only on the total energy spectrum. This peak allows verifying that the gain does not change all along the experiment. This peak disappears when the coincidence mode is required because photoelectrons are not seen in the coincidence mode.

Due to the saturation of the flash ADC and the charge sensitive preamplifier, Alpha particles coming from natural radioactivity induce peaks of saturation in the preamplifier and in the ADC close to the maximum channel (i.e. 3300).

The plot shown in figure 6 highlights the discrimination of the μ-TPC between gammas and neutrons during the neutron energy measurement. In this way the data could be easily filtered to select mainly nuclear recoils before performing the data analysis.

*1) Reconstruction Method*

To measure the neutron energy, the proton energy and the initial proton recoil angle are needed for each event (equation 1). In order to get the 3D tracks, events are rejected of the analysis if the track has a number of time slices (20 ns each) less than three.

To avoid truncated tracks events at less than 4 strips to the edge of the detection zone are removed. This cut enables also to remove events coming from the walls or from the field cage of the detector.

*2) Proton Energy Calculation*

The flash ADC coding the grid signal gives the information to produce the ionization energy spectrum. The energy of nuclear recoils is then calculated event by event by taking into account the ionization energy calibration. The comparison between the ionization spectrum measured and the simulated one by MCNPX and then affected by the ionization quenching factors of H, F and C is shown in fig. 7.

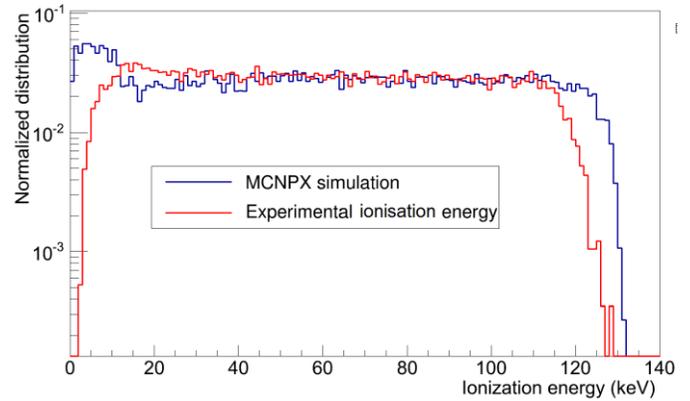

Fig. 7. Ionization energy distribution obtained when a coincidence between strips of pixels X and Y is required (coincidence mode). The X ray calibration has been applied to convert charge, from the signal on the flash ADC, to ionization energy in keV for each event. The blue curve corresponds to the MCNPX simulation affected by the ionization quenching factors of H, F and C.

At energies lower than 15 keV, the distribution decreases rapidly due to the rapid decreasing of quenching factor of carbon and fluorine populating mainly the bump at these energies.

A MCNPX simulation was done to compare the experimental ionization energy distribution to an expected ionization energy distribution. The input neutron field was obtained with TARGET code [20] and is filtered by the solid angle of the system to reach a realistic neutron field, 127 keV with an FWHM of 7.2 keV. Then transport and conversion of neutrons are done in a realistic μ-TPC geometry. The experimental end point, i.e. 123 keV, is 7 keV lower than the simulated one, but remains in the uncertainties obtained by the calibration process.

*3) Initial Proton Recoil Angle Reconstruction*

The sampling of the pixelized anode, every 20 ns, defines the track of the recoil included in the cloud of pixels in three dimensions.

The reconstruction method of the recoil angle includes a fit of a straight line as the major axis of the cloud of pixels in the three spatial dimensions. This method using a linear fit is justified because of the small deviation of proton tracks during their motion at these energies (~100 keV) at low pressure (50 mbar) compared with the size of pixels. This deviation has been calculated by the Monte Carlo code SRIM [18]. The direction vector of the fitted line enables to calculate the angle between the track (fitted line) and the neutron incident direction. As the active area of the detector is wide (116 cm²) and close to the neutron source (72.5 cm), all neutrons are not parallel to the Z-axis. The neutron direction vector is calculated by linking the neutron source position (the target) and the initial proton recoil position. The X and Y positions of each proton track are calculated via the barycenter of each time slice of pixels. The Z position is unknown but it was fixed at the middle of the detection volume. A study with MCNPX simulations has shown that this hypothesis modifies less than 0.01% the mean neutron energy and increases the standard deviation by 6.3%. A scalar product between the unit direction vectors of the neutron and the proton gives the proton initial recoil angle.

To validate the reconstruction method a simulation of the detector was performed. The model uses MAGBOLTZ and SRIM calculations and is based on the model described in [21]. This simulation showed that this method induces a bias lower than 4.5° on the angle due to the diffusion of electrons collected to the grid.

*4) Reconstruction of the Neutron Energy*

Since the ionization energy was measured, the proton IQF can be applied to calculate the proton energy. Once the proton energy and the initial proton recoil angle are measured event by event the neutron energy may be reconstructed via the equation 1. The figure 8 shows the agreement between the data and the equation 1, supposing every event is a proton recoil. This figure 8 shows the interest and the limit of this approximation based on the fact that the hydrogen is the main component of the gas and its neutron elastic scattering cross section is, at these energies, at least 4 times greater than the other components of the gas.

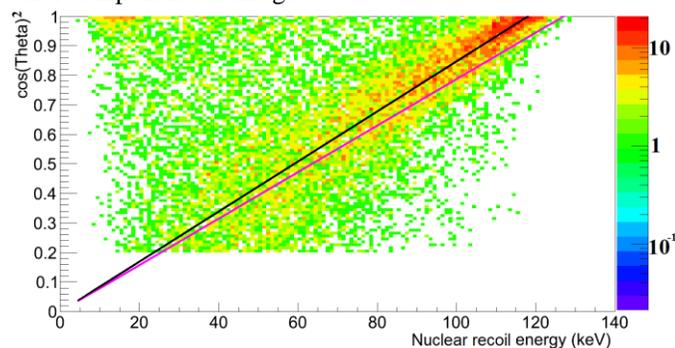

Fig. 8. Square of the cosine of the reconstructed recoil angle versus the nuclear recoil energy. The black and the pink curves correspond to the equation 1 for a proton recoil and respectively a neutron energy of 120 keV and 127 keV. The points represents the data obtained with a neutron field of an expected energy of 127 keV.

The main population of the distribution plotted in the figure 8 follows the equation 1 calculated for a proton recoil and a neutron energy of 120 keV (black curve). The expected neutron energy is 127 keV. The difference comes mainly from the extrapolation of the calibration up to 127 keV and the probable overestimation of the IQF by SRIM.

From the figure 8, we can conclude that:
i) Heavy nuclear recoils (i.e. carbon and fluorine) are obviously wrongly reconstructed because their atomic masses and their IQF are supposed, to produce this plot, equal to the proton ones. These recoils are located on the upper left side of the figure 8.
ii) Protons with an initial recoil angle higher than 40 degrees have energies less than 75 keV. The track reconstruction algorithm is then much less accurate to get the right angle for elastic scattering angles higher than 40 degrees.

The reconstruction method is then more reliable for events with angles lower than 40 degrees. Events with an initial recoil angle higher than 40 degrees are removed from the measurement of the neutron energies. The neutron energy distribution obtained with this cut is shown on the figure 9 (red curve).

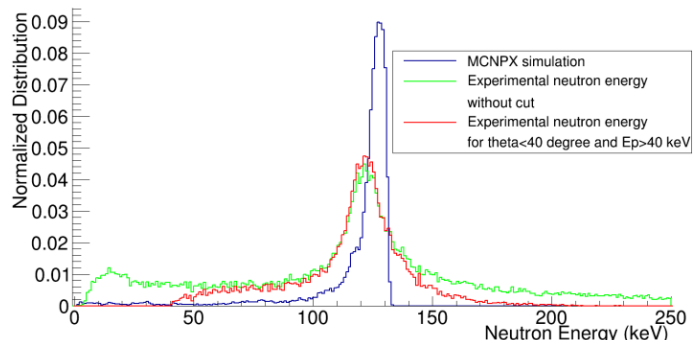

Fig. 9. Experimental neutron energy distributions, obtained by reconstruction, for all events (green curve) or events with recoil angles lower than 40° and nuclear recoil energies higher than 40 keV (red curve). The neutron energy distribution simulated at 127.5 keV with MCNPX is also plotted (blue curve). The difference between the maximum of the experimental and simulated distributions is mainly due to the overestimated IQF given by SRIM and the extrapolation of the linear calibration.

Each distribution was normalized by the integral of the peak between 107 keV and 147 keV. This normalization allows comparing simulation and experimental data.

The reconstructed neutron energy distribution plotted without cuts shows a peak and a little bump. The small bump at low energy is due to heavy recoils which have low energy due to their mass and their IQF. The peak is the one expected at 127 keV. This background is removed for high energies thanks to the cut on the angle (red curve).

Each distribution was fitted by a Gaussian function. The mean of the experimental distribution is 122 keV while the mean of the simulated distribution is 127.5 keV. The figure 8 has already shown this difference. The resolution of the simulated distribution is calculated by the TARGET code with the target thickness and the kinematic calculations. The resolutions, defined as the FWHM over the mean energy, of the simulated and experimental distributions are respectively 8% and 15%.

The neutron energy was measured thanks to this reconstruction method. The angular straggling or the length of the track versus the particle energy are discriminating variables to remove heavy nuclear recoils from the analysis and the tracks of photoelectrons are one order of magnitude longer than the track of nuclear recoils at the same energy allowing their discrimination [21].

*5) Validation of the Reconstruction Algorithm*

In order to validate the reconstruction of the neutron energy distribution, the reconstruction algorithm was tested with simulations of the experimental set up response.

The TARGET code allowed the calculation of the theoretical neutron field, produced by AMANDE at a mean energy of 127.5 keV and a resolution (FWHM/mean) of 8%. Using the MCNPX code, a simulation was performed to obtain simulated proton tracks, induced in the detection volume by the theoretical neutron field. The simulated set up was reproduced as close as possible to the experimental one in order to take into account neutron scattering. Then the detection process of proton tracks was simulated, using MAGBOLTZ and SRIM calculations, to get finally clouds of pixels similar to experimental data. This simulation was based on physical models previously described [21]. The neutron energy distribution was reconstructed with the same algorithm as the one used for experimental data.

The comparison of the experimental (blue curve) and simulated (red curve) neutron energy distributions is shown in the figure 10.

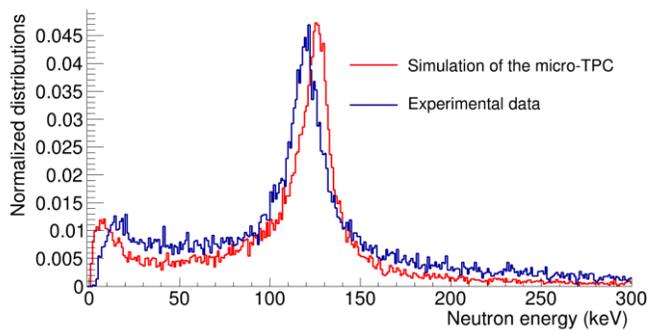

Fig. 10. Simulated and experimental neutron energy distributions obtained with the reconstruction algorithm. The distributions are normalized by their integral between 107 keV and 147 keV. A bias of 4.1 keV is still observed between both maximum of the neutron energy distributions. The theoretical neutron energy distribution has a mean neutron energy of 127.5 keV.

The simulation does not take into account any energy threshold of the µ-TPC and any the heavy ion quenching factor, which explains the difference between both bumps at low energies. The mean of the Gaussian function, fitted to the simulated distribution, is 126.1 keV and the resolution (FWHM / mean) is 12%.

The mean energy of the simulated distribution, reconstructed by the algorithm, is 1.1% lower than the mean of the theoretical energy distribution, used as input. Another estimator of the neutron energy should be used to remove this bias because the theoretical distribution is not a Gaussian function. The resolution of the simulated distribution is 4% wider than the theoretical energy resolution.

The difference between both experimental and simulated mean neutron energies, 4.1 keV, reinforces the idea that the bias on the experimental neutron energy would come mainly from the calibration and the estimation of the quenching factor. The experimental resolution is only 3% wider than the simulated one but the FWHM are the same. Thus the µ-TPC response is well reproduced by the simulations.

## V. CONCLUSION

This experimental campaign demonstrates the ability of our detection system to reconstruct the energy of a mono-energetic neutron field at 127 keV. These measurements have shown the high discrimination of the detector between the photoelectrons and the nuclear recoils: more than 99% of photoelectrons are removed in the coincidence mode.

The mean neutron energy found was 122 keV with an uncertainty of 12 keV mainly due to the calibration and IQF uncertainties.

The resolution (FWHM) of the experimental energy distribution is 15%, while the simulated theoretical resolution is 8%. The reconstruction method will be improved by changing the fitting algorithms to take into account events with energies lower than 40 keV. In the next few months this method will be entirely characterized by the simulation of the detector. Additionally a new chamber will be constructed to reduce the neutron scattering in the walls.